\documentclass[journal]{rmaa}

\title{Physical parameters of three field RR Lyrae stars\altaffilmark{1}}

\author{ A. Arellano Ferro,\altaffilmark{2}
	J. H. Pe\~na,\altaffilmark{2}
        R. Figuera Jaimes,\altaffilmark{3,4}}

\altaffiltext{1}{Based on observations collected at the San Pedro M\'artir Observatory, Mexico}
\altaffiltext{2}{Instituto de Astronom\'ia, Universidad Nacional Aut\'onoma de M\'exico.}
\altaffiltext{3}{European Southern Observatory, Karl-Schwarzschild Stra$\beta$e 2,
85748 Garching bei M\"unchen, Germany}
\altaffiltext{4}{SUPA, School of Physics and Astronomy, University of St.
Andrews, North Haugh, St Andrews, KY16 9SS, United Kingdom}

\fulladdresses{
\item A. Arellano Ferro, J. H. Pe\~na:
Instituto de Astronom\'ia, Universidad Nacional Aut\'onoma de
M\'exico, Apdo. Postal 70-264, M\'exico D. F. CP 04510, M\'exico 
(armando@astroscu.unam.mx; jhpena@astroscu.unam.mx).

\item R. Figuera Jaimes:
European Southern Observatory, Karl-Schwarzschild Straße 2, 85748
Garching bei Munchen, Germany (rfiguera@eso.org, robertofiguera@gmail.com)}

\listofauthors{A. Arellano Ferro, J. H. Pe\~na \& R. Figuera Jaimes}
\indexauthor{Arellano Ferro, A.}
\indexauthor{Pe\~na, J. H.}
\indexauthor{Figuera Jaimes, R.}

\abstract{Str\"omgren $uvby-\beta$ photometry of the stars classified as RR Lyrae stars
RU Piscium, SS Piscium and TU Ursae Majoris has been used to estimate their iron
abundance,
temperature, gravity and absolute magnitude. The stability of the pulsating period is
discussed. The nature of SS Psc as a RRc or a HADS is addressed. The reddening of
each star is estimated from the Str\"omgren colour indices and reddening sky maps. The
results of three approaches to the determination of [Fe/H], $T_{\rm eff}$ and
$\log(g)$ are discussed: Fourier light curve decomposition, the Preston $\Delta S$
index and the theoretical grids on the $(b-y)_o - c_1{_o}$ plane.}

\resumen{La fotometr\'ia Str\"omgren $uvby-\beta$ de las estrellas clasificadas como
RR
Lyrae RU Piscium, SS Piscium and TU Ursae Majoris, ha sido empleada para estimar
la abundancia
de hierro, la temperatura efectiva, la gravedad superficial y la magnitud absoluta.
Se discute la variabilidad secular del per\'iodo de pulsaci\'on. El enrojecimiento de
cada estrella se ha estimado a partir de los colores de Str\"omgren y de mapas
celestes de enrojecimiento. Se reportan y comparan los resultados para [Fe/H], $T_{\rm
eff}$ y $\log(g)$ obtenidos por alguno de los siguientes m\'etodos: Descomposici\'on
de Fourier de la curva de luz, el \'indice $\Delta S$ de Preston y la comparaci\'on
con las mallas te\'oricas en el plano $(b-y)_o - c_1{_o}$.} 

\keywords{Variable Stars-RR Lyrae, photometry, $uvby-\beta$ photometry }

\shortauthor{Arellano Ferro et al.} 
\shorttitle{Three field RR Lyrae stars}

\begin{document}

\maketitle

\section{Introduction}
RR Lyrae stars (RRL) are radial pulsators that can be active in the fundamental mode
(RRab), first overtone (RRc) or both (RRd). They are common in globular clusters where
they populate the horizontal branch (HB). Amplitude and period modulations discovered
early in the XX$^{th}$ century (Bla$\check{\rm z}$ko 1907) have now being identified
in a large
number of RRab and RRc stars.  Perhaps the most prominent of the RRL properties is the
correlation between their pulsation period, luminosity and [Fe/H] content: RRL
in more metal poor clusters are brighter and have longer periods, the Oosterhoff type
II clusters, than in their counterparts Oosterhoff type I (Oosterhoff 1939).
The $M_V$-[Fe/H] relationship has been calibrated by several authors (Chaboyer 1999;
Cacciari \& Clementini 2003; Arellano Ferro et al. 2008) and confirms the RRL stars as
good distance indicators.

Long time series of RRL stars produced by space missions have shown to be very
informative on many of the most detailed pulsational properties of these stars, e.g.
change of period modes, Bla$\check{\rm z}$ko periods, period doubling, presence of
non-radial modes,
etc. An outstanding example is the analysis of $Kepler$ space mission data for
the star V445 Lyr recently carried out by Guggenberger et al. (2012). CCD
time series of globular clusters have also proven very useful in identifying RRL
variables and in estimating their physical parameters such as luminosity and
metallicity from the Fourier decomposition of their light curves (e.g. Arellano Ferro
et
al. 2011 and references therein).

Determination of physical parameters of RRL stars in globular clusters, despite the
fact that they are faint, has the advantage that high quality photometry can
now be achieved with the use of Differential Imaging Approach techniques (e.g. Alard
2000; Bramich 2008). For the brighter isolated field RRL's long time series are rarer
and hence their physical parameters and precise position on the HB are more 
difficult to determine. 

Fourier decomposition of light curves of seven field RRL in Bootes and a
discussion of their physical parameters from Str\"omgren photometry was carried out
by Pe\~na, et al. (2009).  Very recently Pe\~na et al. (2012) have published
Str\"omgren data for three RRL's: RU Psc, SS Psc and TU UMa. In the present paper
we discuss the use of those data to calculate some physical parameters of these stars.

The paper is organized as follows: In $\S$ 2 we describe the data. In $\S$ 3 the
fundamental characteristics of the three sample stars are presented and the
secular period variatons are discussed. In $\S$ 4 the Fourier light
curves decomposition is performed and the implied physical parameters are given. In
$\S$ 5 we calculate the reddening. In $\S$ 6 we present the iron abundance estimation
from the Preston $\Delta S$ parameter. In $\S$ 7 the mean values and pulsating ranges
of $T_{\rm eff}$ and $log(g)$ are estimated by comparing with thoretical grids. In
$\S$ 8 we summarize our conclusions.

\begin{figure}[t]
\includegraphics[width=8.1 cm,height=8.0 cm]{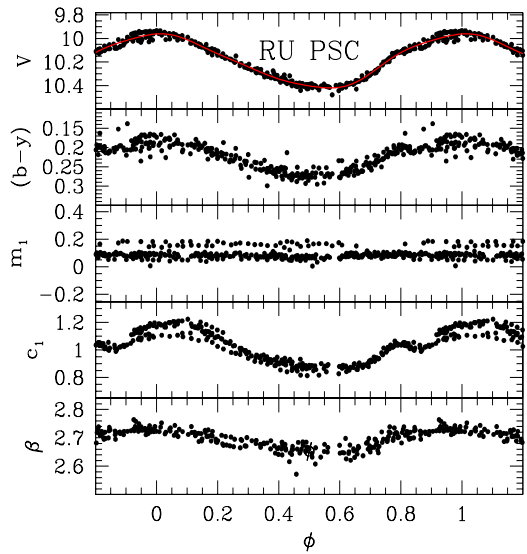}
\caption{Light and colour curves of RU Psc phased with the period 0.391156d. The red
line is a fit of the form of eq. \ref{eq:fourier} with four harmonics.}
\label{fig:RU}
\end{figure}

\section{Data}
\label{sec:Data}

All observations were carried with a multichannel spectrophotometer mounted on the
1.5 m telescope  of the Observatorio Astron\'omico Nacional in San Pedro M\'artir
(OAN-SPM), Mexico. This spectrophotometer simultaneously observes
the Str\"omgren $uvby$ bands and almost simultaneously the two filters
that define H$\beta$ (Schuster \& Nissen 1988). The observations were obtained on 9
nights between 1989 and 1995 for Ru Psc for a total of 310 individual measurements.
For SS Psc the observations are from 10 nights spanning from
1992 to 1995 and a total of 139 measurements. For Tu UMa the data are from 14 nights
between 2004 and 2009 and a total of 50 independent observations.

\begin{figure}
\includegraphics[width=7.8 cm,height=7.6 cm]{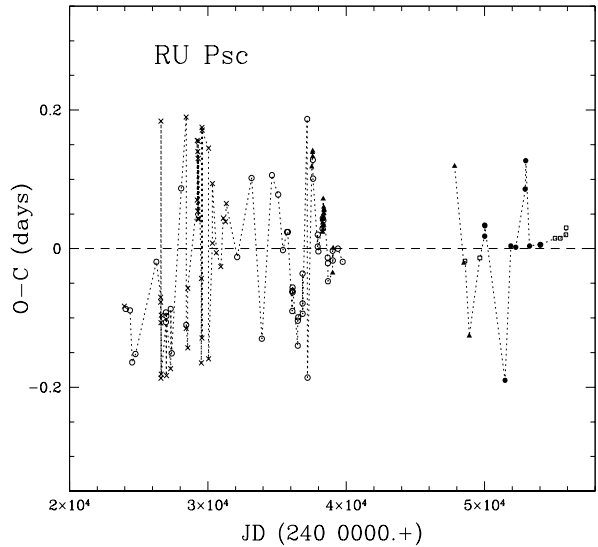}
\caption{O-C residuals of RU Psc calculated with the ephemerides 2440143.40270 +
0.390385~E spanning from 1924 to 2011. Symbols are; open circles: unknown; crosses:
photographic; open squares: visual; filled triangles: photoelectric; filled circles:
CCD. {\bf Error bars for photoelectric and CCD data would be of the size of the
symbols while for visual observations they may reach about 0.02d}.}
\label{fig:OC}
\end{figure}

\section{Sample stars}
\label{sec:sample}

{\bf RU Psc}. This is an RRc star with a variable pulsation
period that ranges between 0.390318 and 0.3900421 according Mendes de Oliveira \&
Nemec
(1988). The period listed by Kholopov et al. (1987) is 0.390385d. In Fig. \ref{fig:RU}
light and colour curves are phased with the best period found in our data 0.391156d
$\pm$ 0.000002d, which is much larger even than the largest period found by
Mendes de Oliveira \& Nemec (1988). As has been found by previous authors, the
presence of nightly shifts that cannot be taken into account by a single period is
clear. A similar behaviour has been seen in double mode RRd stars. Also the
cycle-to-cycle variations were attributed to the
Bla$\check{\rm z}$ko effect with a period of about 28.8d according to Tremko (1964).
With an
analysis of more than 1100 photometric observations, Mendes de Oliveira \& Nemec
(1988) concluded that RU Psc is not a double mode or RRd star but that the
peculiarities seen in the photometry of RU Psc are due primarily to rapid and
irregular period changes, possibly of the Bla$\check{\rm z}$ko nature, but they also
ruled out the
presence of a 28.8d Bla$\check{\rm z}$ko period.

\begin{figure}[t]
\includegraphics[width=8.1 cm,height=8.0 cm]{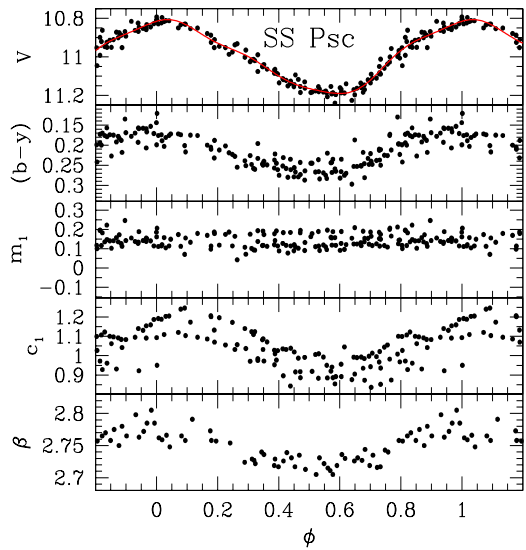}
\caption{The light and colour curves phased with the period 0.288712d.
The red solid curve is the fit with four harmonics in eq. \ref{eq:fourier}.}
\label{fig:SS}
\end{figure}

Given the variable nature of the period, probably a more appropriate insight can be
found by examining the O-C residuals from a given ephemerides.
From our observations we determined two new times of maximum $V$ light at
HJD 2447835.6660d and HJD 2448896.8778d.
In the RRL data base of GEOS (2012) (http://rr-lyr.ast.obs-mip.fr)
one can find a list of 122 times of maximum spanning from 1924 to 2011.
The O-C residuals were calculated with the ephemerides 2440143.40270 + 0.390385~E
of Kholopov et al. (1987) and they are plotted in Fig. \ref{fig:OC}. 
{\bf The errors in the O-C values from photometric or CCD data are better that a few
thousands of a day (e.g. Agerer \& H\"ubscher 1996) whereas the visual estimates can
reach an uncertainty of about 0.02d. Thus the irregular O-C oscilations seen in Fig.
\ref {fig:OC} are real and not a consequence of the uncertainties in the time of
maximum estimations.} If the time of maximum light deviations about a mean are
produced by
the presence of a secondary oscillation, its frequency should be detected in a period
analysis of the residuals, after removing the primary frequency, which is naturally
contained in the O-C residuals via the ephemerides.
We have used the {\tt period04} program (Lenz \& Breger 2004) to search for
periodicities in the residuals. If the star was an RRd, a period of $\sim$0.527d would
be expected. We found no significant period in the vicinity of the above period, hence
we follow Mendes de Oliveira \& Nemec (1988) in concluding that RU Psc is not an RRd
star. Most recently a group of double mode RRL stars with a period ratio of $P2/P1
\sim$ 0.61 have been identified (e.g  Soszy\'nski et al. 2009, Moskalik et al. 2012).
This has been interpreted as due to the presence of a non-radial mode. The expected
secondary period would be $\sim$0.238d which also was not found.

\begin{figure}[t]
\includegraphics[scale=0.6]{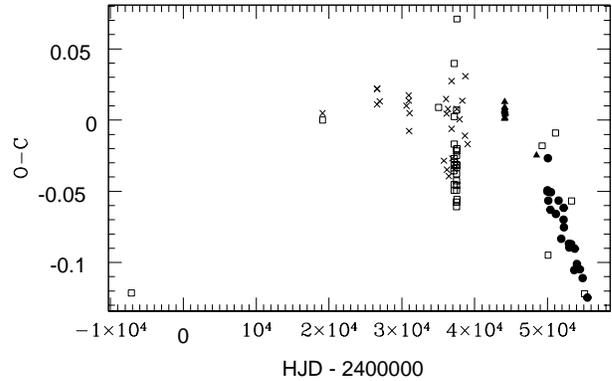}
\caption{O-C diagram for the sample of maxima of SS Psc. Symbols are: crosses:
photographic; squares: visual; filled triangles: photoelectric; filled circles: CCD. 
{\bf Error bars for photoelectric and CCD data would be of the size of the
symbols while for visual observations they may reach about 0.02d}.}
\label{fig:OCDSS}
\end{figure}

The Bla$\check{\rm z}$ko nature of RU Psc is controversial. Bla$\check{\rm z}$ko
variations are now
well known in a large number of RRc stars (e.g. in M55 Olech et al. 1999; in the LMC
Alcock et al. 2000; in NGC 6362 Olech et al. 2001; in three RRc stars
in the OGLE database Moskalik \& Poretti 2003; in Omega Centauri Moskalik \& Olech
2008; in M53 Arellano Ferro et al. 2012) however the light curve of RU Psc differs
from most of the known modulated RRc stars in that its modulation is mostly in phase
(period) while in amplitude the modulations, if real, are very mild. This suggests
that the star undergoes an irregular period variation rather than having a real
modulation typical in Bla$\check{\rm z}$ko variables. The light and colour curves of
Fig. \ref{fig:RU}, being formed from data acquired over 6 years, show a dispersion
produced by the period irregular fluctuations.

{\bf SS Psc}. We find this star listed as RRL and as $\delta$ Scuti
star in different sources. The star is listed as RRc in the GCVS (Kholopov et al.
1987) with a period of 0.28779276d. Until 1977 it was included in photometric
programs of RRL stars (e.g. Bookmeyer et al. 1977).
The star is declared as an RRL in the SIMBAD data base.
These facts perhaps led Pe\~na et al. (2012) to include the star in their program 
of field RRL $uvby$ photometry.

The star was the subject of a $uvby$ photometric study of McNamara \& Redcorn
(1977). These authors concluded, based on its apparently rich metallicity,
that the star is not an RRc but a dwarf Cepheid (or RRs) which are considered
more massive (1-2 $M_{\odot}$) post main sequence stars than the less massive (0.5-0.7
$M_{\odot}$) post giant branch stars.

SS Psc is included in the sample of Antonello et al. (1986) among high amplitude
$\delta$ Scuti stars (HADS). At the Fourier decomposition of the light curves of their
sample stars, Antonello et al. note that among the low harmonic amplitude
ratio $R_{21}=A_2/A_1$ stars, SS Psc is indistinguishable of RRc stars, like other
HADS, and that on the amplitudes plane $A_2$ vs $A_1$ the star seems a prolongation
of the RRc domain toward lower amplitudes. They conclude that SS Psc may be a link
between the two classes. The star is included in the list of $\delta$ Scuti star of
Rodr\'iguez et al. (1994).  

The mean gravity calculated McNamara \& Redcorn (1977), $3.29\pm0.1$, is only
marginally larger than in RRc stars ($\sim 3.1$) and it is based on only 18
photometric data obtained on a single night. Similarly the  metallicity of the star
has been estimated as such by the rich metallic-line appearance of the spectrum, but
to our knowledge no detailed abundance analysis has been performed. In our opinion
insufficient arguments have been given to classify SS Psc either as a
HADS or as RRc, and a dedicated spectroscopic detailed study is highly desirable.
{\bf In section \ref{sec:fourier} we shall further discuss the nature of SS Psc from
the Fourier decomposition of the $V$ light curve in Fig. \ref{fig:SS}}.

Our data of SS Psc are 139 data points in a time span of 1090 days between JD 2448891
and 2449981. The phasing with Kholopov et al.'s (1987) period of 0.28779276d gave a
small displacements of the maximum light. A PDM analysis of our data produced a period
of (0.288712 $\pm$ 0.000005)d.

In the RRL data base of GEOS (2012) (http://rr-lyr.ast.obs-mip.fr) one can find a
list of 96 times of maxima and the corresponding O-C diagram built with the
ephemerides
2419130.305 + 0.28779276 E is reproduced in Fig. \ref{fig:OCDSS}. {\bf Visual
observations show a particularly large scatter at HJD $\sim$2443800 which may be due
to
their larger uncertainty. We have included them for completeness but if they are
ignored the O-C diagram strongly suggests} an abrupt period change sometime around HJD
2445000. However it could be
argued that the O-C is not inconsistent with a secularly decreasing period if older
visual observations are considered. In this case however, the
time base would be too short for a clear difinition of the period change.
We feel that only further observations can disentangle the above possibilities.

{\bf TU UMa}. This is a RRab star with a $V$ amplitude 
of about 0.9 mag. and a period of 0.5576587d (Kholopov et al. 1987). After the
work of Szeidl, Olah \& Mizser (1986) the star has been well known for undergoing
secular
period variations and some cyclic variations of the O-C residuals  with a period of
about 23 years, which led these authors to suggest that the star is in a binary
system.
Analysis of the times of maximum light have been used by several authors to propose 
a model which is a combination of a quadratic pulsation ephemerides (a secularly
decreasing period) and an orbital timing model which fits the O-C data within the
photometry uncertainties (Saha \& White 1990; Kiss et al. 1995;  Wade et al. 1999).

Wade et al. (1999) also show convincing evidence that the center-of-mass velocities
for TU UMa are consistent with an orbital solution with period of about 23 yr, an
eccentricity of 0.79 and a mass for the companion star of $\sim 0.4~M_\odot$.

\begin{figure}[t]
\includegraphics[width=7.8 cm,height=6.0 cm]{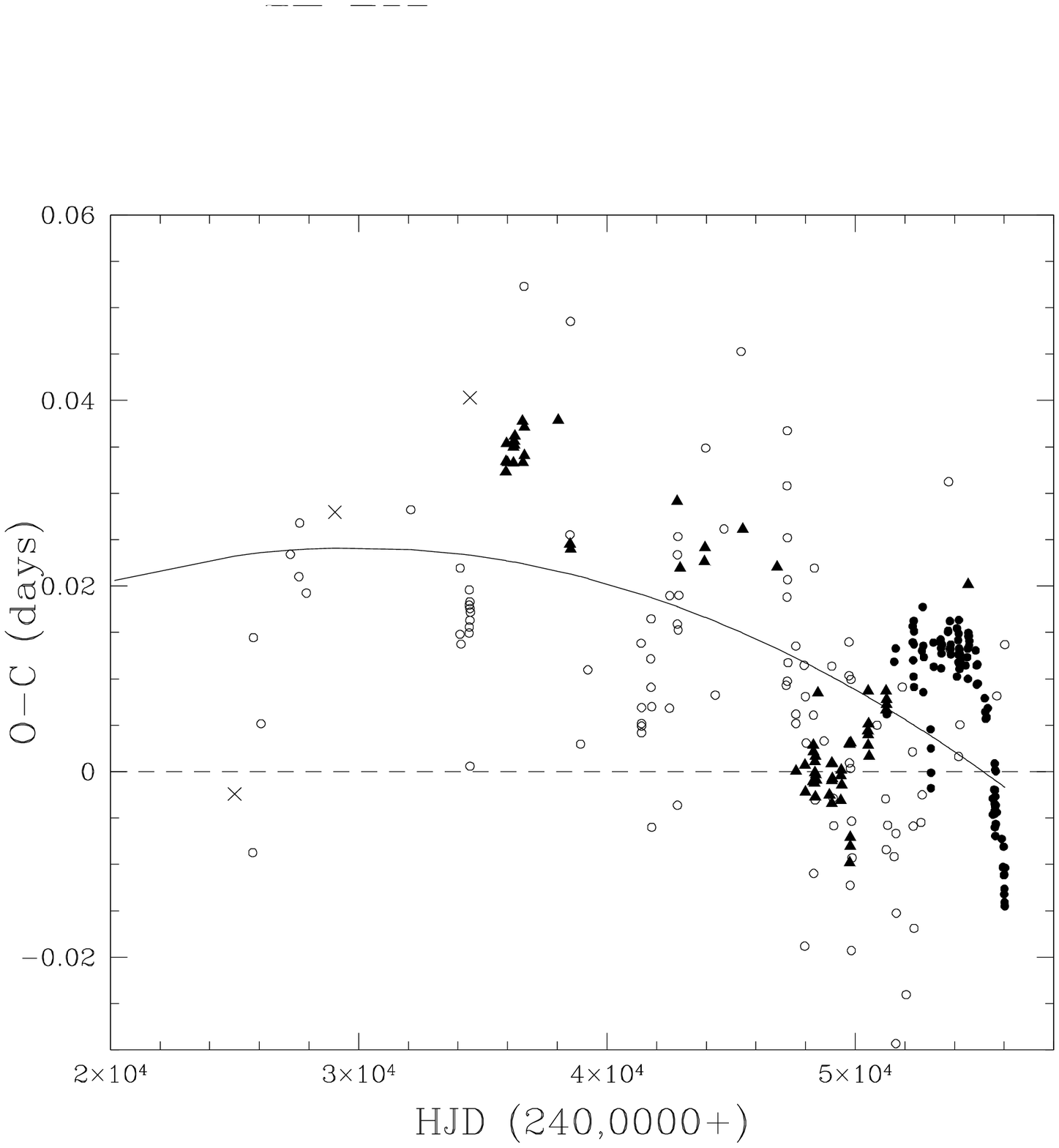}
\caption{O-C diagram for the sample of maxima of TU UMa. Symbol are; crosses:
photographic; open circles: visual; filled triangles: photoelectric; filled circles:
CCD. {\bf The O-C uncertainties are as in Figs. \ref{fig:OC} and \ref{fig:OCDSS}}. The
black solid curve is the parabolic fit to all data.}
\label{fig:TUUMa}
\end{figure}

\begin{figure}[t]
\includegraphics[width=8.1 cm,height=8.0 cm]{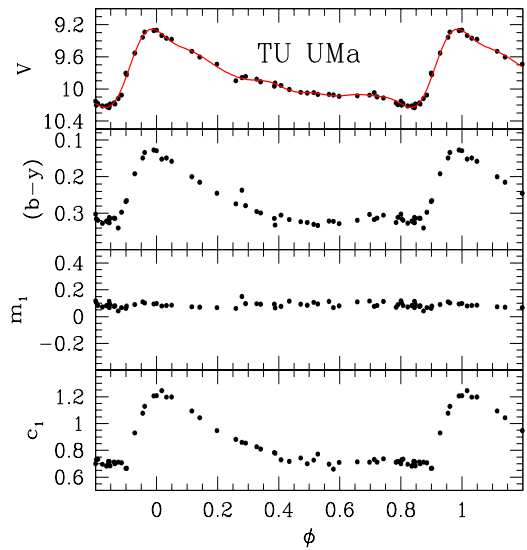}
\caption{Light and colour curves of TU UMa phased with the ephemerides 2454552.90345
+0.5576587 E. The red solid curve is the fit with nine harmonics in eq.
\ref{eq:fourier}. H$\beta$ data are not available in Pe\~na et al. (2012) for this
star.}
\label{fig:TU}
\end{figure}

While cyclic period variations have been reported in several RRL, the
nature of their origin as being caused by multiperiodicity, light time effects
or hydromagnetism is still under debate (see Derekas et al. (2004) for a
discussion). The continuous observations of both field and cluster RRL stars, the
registration of epochs of maximum light and the analysis of line profile variation on
high resolution spectra will, however, provide the basis to further interpret the
period variations.

A large compilation of times of maximum light is found in the GEOS (2012)
(http://rr-lyr.ast.obs-mip.fr) data base. We have noticed that this compilation
misses 33 of the 41 times of maximum listed in Table 2 of Wade et al. (1999). In Fig.
\ref{fig:TUUMa} we present the complete O-C residuals, which for a direct comparison
with the models of Wade et al. (1999), were calculated with the ephemerides used by
these authors 2425760.4364 + 0.557658109 E, supplemented with the time of maximum
HJD 2454552.90345 calculated in this work.

The O-C clearly shows the secular variation of the pulsating period which can be
represented by

\begin{equation}
\label{eq:fit}
    O-C=0.02351+1.71274 \times 10^{-7}  E - 1.17113\times 10^{-11} E^2\nonumber
\end{equation}

\noindent
this implies a period decreasing rate of -1.3 ms/yr. Other determinations range from
$-9.9$ ms/yr (Kiss et al. 1995) to $-1.7$ ms/yr (for some models of Wade et al. 1999).
Since the model calculation by Wade et al. (1999) numerous new O-C are available,
mostly from CCD measurements. Inclusion of these data in the O-C diagram (Fig.
\ref{fig:TUUMa} for HJD $\geq$ 2451545.) neatly delineate another 'orbital loop'.
We feel it is worth waiting for another ten years of accurate times of maximum light
registration before the orbit is accurately solved.

Our data consists of 50 data points obtained between JD 2453102 and 2454881. 
Fig. \ref{fig:TU} shows the light and colour curves phased with the period of Kholopov
et al. (1987), which we shall use for the Fourier decomposition.

\section{Fourier Analysis} 
\label{sec:fourier}

Given the light curves of the RRL stars we make an attempt to estimate their
iron abundance, effective temperature and absolute magnitude using the Fourier light
curve decomposition approach. For SS Psc, given its unclear nature as an RRL (see
section \ref {sec:sample} and further below in this section for a discussion), we
shall not apply this approach since the involved calibrations may not be applicable.

The light curve of a periodic variable can be represented by an equation of the form:

\begin{equation}
\label{eq:fourier}
    m(t)=A_0 + \sum_{k=1}^N A_k \cos(\frac{2\pi}{P}k
    (t-E)+\phi_k),\nonumber
\end{equation}

\noindent
and the phase and amplitude ratio Fourier parameters are defined as
$\phi_{ij} =  j\phi_i - i\phi_j$ and $R_{ij} =  A_i / A_j$.

Decomposing the light curve in its harmonics has proved to be useful in estimating
physical parameters in field RRL stars (e.g. Pe\~na et al. 2009) and in globular
clusters (e.g. Arellano Ferro et al. 2011 and references therein), through the use of
semi-empirical relationships (e.g. Jurcsik \& Kov\'acs 1996; Morgan, Wahl \&
Wieckhorst 2007).

Epochs and periods used in the application of eq. \ref{eq:fourier} and the resulting 
amplitudes $A_i$ and phases $\phi_{ij}$ for each harmonic for the three sample stars
are reported in Table \ref{tab:ampdisp}.

\begin{table*}[!t]\centering
{\small
\setlength{\tabnotewidth}{\columnwidth}
  \tablecols{13}
 \setlength{\tabcolsep}{1.0\tabcolsep}
 \caption{Fourier coefficients derived from the fit}
\label {tab:ampdisp}
  \begin{tabular}{lccccccccccc}
 \hline
ID     &    $E_0$   & P          &  $A_0$  & $A_1$ &  $A_2$& $A_3$ & $A_4$ & $\phi_{21}$ & $\phi_{31}$ & $\phi_{41}$ & N \\
       &$2400000.0+$& (days)     &         &       &       &       &       &             &             &                 \\
\hline
RU Psc & 47835.6687 & 0.391156   & 10.192  & 0.224 & 0.020 & 0.009 & 0.010 & 4.553   
&4.518  & 2.757  & 4 \\
       &            &            &  0.001  & 0.002 & 0.002 & 0.002 & 0.002 &    0.084
  &   0.178    &    0.171    &   \\
SS Psc & 49980.7713 & 0.288712   & 11.001  & 0.181 & 0.020 & 0.006 & 0.010 &  4.972 &
5.100 & 3.025& 4 \\
       &            &            &  0.002  & 0.003 & 0.003 & 0.003 & 0.003 &    0.155
  &   0.499    &    0.309    &   \\
TU UMa & 54552.9035 & 0.5576587  &  9.857  & 0.328 & 0.164 & 0.115 & 0.071 &    3.893 
  &    8.150    &    6.265    & 7 \\
       &            &            &  0.005  & 0.007 & 0.007 & 0.007 & 0.007 &    0.060    &    0.090    &    0.130    &   \\
\hline

  \end{tabular}
 }
\end{table*}

\begin{table*}
\centering
\setlength{\tabnotewidth}{\columnwidth}
  \tablecols{9}
 \setlength{\tabcolsep}{1.0\tabcolsep}
 \caption{Fourier decomposition physical parameters for two RR Lyrae stars}
\label{tab:Physpar}
  \begin{tabular}{lcccccc}
 \hline
ID       & [Fe/H]$_{\rm ZW}$ & $\log(T_{\rm eff})$  & $M_V$ & $\log(L/L_\odot)$
& $\mu_0$ &   distance (pc) \\
\hline
RU Psc   &   $-1.62\pm 0.14
$&$3.859\pm0.003$&$0.51\pm0.04$&1.724&$9.65\pm0.04$&$851\pm16$\\
TU UMa   &   $-1.53\pm 0.14
$&$3.809\pm0.003$&$0.60\pm0.04$&1.672&$9.25\pm0.04$&$710\pm15$\\
\hline
  \end{tabular}
\end{table*}

For the RRab star, TU UMa, we used the calibration of Jurcsik \& Kov\'acs (1996):
 
\begin{equation}
\label{eq:JK}
	{\rm [Fe/H]}_{\rm J} = -5.038 ~-~ 5.394~P ~+~ 1.345~\phi^{(s)}_{31},
\end{equation}

\noindent
{\bf The standard deviation of this calibration is 0.14 dex (Jurcsik 1998) and} 
surrenders [Fe/H]$_{\rm J}$ in a scale that can be transformed into the
Zinn \& West (1984) scale [Fe/H]$_{\rm ZW}$ through the relationship [Fe/H]$_{\rm J}$
=
1.43 [Fe/H]$_{\rm ZW}$ + 0.88 (Jurcsik 1995). The above equation can be used if the
light
curve satisfies the {\it compatibility parameter} $D_m$ which should be smaller than
3.0 (Jurcsik \& Kov\'acs 1996). For TU UMa we calculated $D_m = 2.1$.

The effective temperature was estimated from the calibrations of Jurcsik (1998)

\begin{equation}
	\log(T_{\rm eff})= 3.9291 ~-~ 0.1112~(V - K)_o ~-~ 0.0032~{\rm [Fe/H]},
\end{equation}

\noindent
with 

$$ (V - K)_o= 1.585 ~+~ 1.257~P ~-~ 0.273~A_1 ~-~ 0.234~\phi^{(s)}_{31} ~+~ $$
\begin{equation}
~~~~~~~ ~+~ 0.062~\phi^{(s)}_{41}.
\end{equation}

{\bf Equation (4) has a standard deviation of 0.0018 (Jurcsik 1998), but
the accuracy of $\log(T_{\rm eff})$  is mostly set by the colour equation (5).
The error estimate on $\log(T_{\rm eff})$ is 0.003 (Jurcsik 1998).}
 
The value of the absolute magnitude comes from the calibration of Kov\'acs \& Walker (2001);

\begin{equation}
\label{eq:KWab}
M_V(K) = ~-1.876~\log(P) ~-1.158~A_1 ~+0.821~A_3 +K.
\end{equation}

\noindent 
{\bf which has an standard deviation of 0.04 mag.} The value of $K = 0.41$ was adopted
to scale the luminosities of RRab stars with
the distance modulus of 18.5 for the Large Magellanic Cloud (LMC) (see the discussion
in Arellano Ferro et al. 2010, in their section 4.2).

For the RRc star RU Psc we calculated [Fe/H]$_{\rm ZW}$ from the calibration of 
Morgan et al. (2007);

\begin{eqnarray}
\label{eq:MORG}
{\rm [Fe/H]}_{\rm {\rm ZW}} = 52.466 P^2 - 30.075 P + 0.131 (\phi^{(c)}_{31})^2
\nonumber \\
+ 0.982 \phi^{(c)}_{31} -4.198 \phi^{(c)}_{31} P + 2.424,
\end{eqnarray}

\noindent
{\bf This calibration provides iron abundances with a standard deviation
of 0.14 dex.} $T_{\rm eff}$ comes from the calibration of Simon \& Clement (1993);

\begin{equation}
\label{eq:SC93}
	\log(T_{\rm eff}) = 3.7746 ~-~ 0.1452~\log(P) ~+~ 0.0056~\phi^{(c)}_{31},
\end{equation}

\noindent
and $M_V$ is obtained from the calibration of Kov\'acs (1998);

\begin{equation}
\label{eq:KWc}
M_V(K) = ~-0.961~P ~-0.044~\phi^{(s)}_{21} ~+4.447~A_4 + 1.061.
\end{equation}

\noindent
{\bf with an error of 0.042 mag.}

Following Cacciari et al. (2005), we have used eq. \ref{eq:KWc} with the zero point 
1.061 in order to bring the absolute magnitudes in agreement with the mean magnitude
for the RRL stars in the LMC, $V_0 = 19.064 \pm 0.064$ (Clementini et al. 2003).

The results obtained from this analysis are compiled in Table \ref{tab:Physpar}. The
absolute magnitude $M_V$ was converted into $\log(L/L_\odot)$ using $M_{bol}{\odot}$
= +4.75 and the colour excesses given in $\S$ 5.

At this point it is worth mentioning that, although calibrations exist 
to estimate the stellar mass from the Fourier parameters for the RRc (Simon \& Clement
1993) and the RRab (Jurcsik 1998) stars, Cacciari et al. (2005) have shown that the
results are unreliable and therefore we have refrained from using such calculations.

\begin{table*}[!t]
\centering
\setlength{\tabnotewidth}{\columnwidth}
  \tablecols{5}
 \setlength{\tabcolsep}{1.0\tabcolsep}
 \caption{$\Delta S$ for the sample stars.}
\label{tab:ephTAB}
  \begin{tabular}{lcccc}
 \hline
ID & Bayley's type & Spectral type & $\Delta S$ \tabnotemark{1}& [Fe/H]($\Delta S$) \\
\hline
RU Psc &  RRc  & A7-F3 & $7.65\pm1.77$ & $-1.61\pm0.27$\\
SS Psc &     & A7-F2 & $2.95\pm1.34$ & $-0.88\pm0.21$ \\
TU UMa &  RRab & F2    & $6.10\pm1.83$ & $-1.37\pm0.29$ \\
\hline
 \tabnotetext{1}{Suntzeff, Kraft \& Kinman (1994)}
 \end{tabular}
\end{table*}

\begin{figure*}
\includegraphics[width=14. cm,height=4.0 cm]{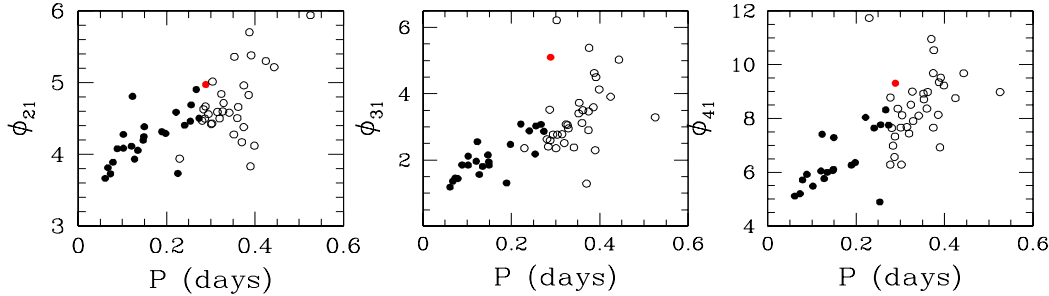}
\caption{Phases  $\phi_{21}$, $\phi_{31}$ and $\phi_{41}$ as function of period for a
sample of HADS and HASXP (filled circles) and for some RRc stars in globular
clusters (open circles). The red point represents SS Psc. See text for discussion.}
\label{fig:PHIs}
\end{figure*}

{\bf The case of SS Psc is worth a separate comment. The coverage and density of the
$V$ light curve of Fig. \ref{fig:SS}
allow the Fourier decomposition in four harmonics with amplitudes and phases listed
in Table \ref{tab:ampdisp}. Given its unclear classification
as a HADS or as a RRc, in Fig \ref{fig:PHIs} we plotted its Fourier phases
$\phi_{21}$, $\phi_{31}$ and $\phi_{41}$ as a function of period for a sample of HADS
and HASXP stars taken from the paper of Antonello et al. (1986) and a sample of
RRc stars in globular clusters taken from several papers published by our group
(Arellano Ferro et al. 2004; 2008; 2011; Bramich et al. 2011; Kains et al. 2012;
L\'azaro et al.
2006). Its $\phi_{21}$ and $\phi_{41}$ phases indicate that SS Psc tends to mix better
with the RRc stars than with the HADS and that $\phi_{31}$ is peculiarly large. Hence
its [Fe/H] and $T_{\rm eff}$ from eqs. \ref{eq:MORG} and \ref{eq:SC93} would be
peculiar and not reliable as for an standard RRc, thus we have do not report here
the values. While these arguments seem to rule out SS Psc as a HADS they indicate that
star remains as peculiar among the RRc star.}

\section{Reddening}
The determination of the reddening in field stars is complex, however, a good
estimation can be obtained from the maps of Galactic dust reddening and extinction of
Schlegel, Finkbeiner \& Davis (1998). These maps give values of E($B-V$)  of 0.047,
0.051, 0.023 in the directions of RU Psc, SS Psc and TU UMa respectively, which
correspond to values of E($b-y$) 0.034, 0.037 and 0.016. 

From $uvby-\beta$ photometry for F type dwarfs and giants one could also use the
calibration of Crawford (1975). We have calculated the magnitude-weighted means of
$b-y$, $m_1$, $c_1$ and $\beta$ from the colour curves in Figs. \ref{fig:RU},
\ref{fig:SS} and \ref{fig:TU}.  For TU UMa there are no $\beta$ measurements, thus
for this star we used the mean values of Hauck \&  Mermilliod (1998).
SS Psc and TU UMa are a bit beyond the validity of the calibration ($\beta$ is too
large), and the calculation implied some extrapolation. TU UMa correctly fits in
the calibration ranges. We found $E(b-y$) 0.014, 0.044 and -0.006 for RU Psc,  SS Psc
and TU UMa respectively. Using their own photometry and the same calibration, McNamara
\& Redcorn (1977) calculated $E(b-y$)=0.016 for SS Psc. 

Based on the above calculations we adopted $E(b-y$) as 0.01, 0.04 and 0.00 for  RU
Psc, SS Psc and TU UMa respectively.

\section{[Fe/H] from the $\Delta S$ parameter}

It is of interest to compare the values of [Fe/H] calculated from the Fourier 
decomposition approach with the estimation via the $\Delta S$ metallicity parameter
defined by Preston (1959) as $\Delta S$ = 10[Sp(H)-Sp(CaII)] that is, the difference
between the hydrogen and K-line types in units of tenths of spectral class. Average
values of $\Delta S$ for the three stars in our sample are reported by Suntzeff, Kraft
\& Kinman (1994) as  $7.65 \pm 1.77$ for RU Psc, $2.95 \pm 1.34$ for SS Psc and $6.10
\pm 1.83$ for TU UMa. Suntzeff, Kraft \& Kinman (1994) also provide the following 
empirical relations; ${\rm [Fe/H]} = -0.155 \Delta S - 0.425$ for RRc stars and ${\rm
[Fe/H]} = -0.158 \Delta S -0.408$ for RRab stars. Hence the corresponding
values of [Fe/H] are  $-1.61 \pm 0.27$, $-0.88 \pm 0.21$, and $-1.37 \pm 0.29$ for RU
Psc, SS Psc and TU UMa respectively. For TU UMa Layden (1994) compiles three values of
$\Delta S$ from the literature that average 6.9 and then [Fe/H]=-1.50.

\begin{table*}[!t]\centering
\setlength{\tabnotewidth}{\columnwidth}
  \tablecols{14}
 \setlength{\tabcolsep}{1.0\tabcolsep}
 \caption{[Fe/H], $T_{\rm eff}$ and $\log(g)$ determinations for the sample stars.}
\label{tab:ranges}
  \begin{tabular}{lcccccc}
 \hline
ID     &[Fe/H]$_{\rm ZW}$ &[Fe/H]($\Delta S$)&[Fe/H]   & $T_{\rm eff}$  &  $T_{\rm
eff}$
range
 & $\log(g)$ range   \\
       &              & &Adopted & Fourier        &          &       \\
\hline
RU Psc &   $-1.62$   &$-1.61$ & $-1.5$  &    7229  & 6200 - 6800 &  2.2-2.7\\
SS Psc &          &$-0.88$ & $-1.0$  &          & 6500 - 7500 & 2.5-3.2\\
TU UMa &   $-1.53$   &$-1.37$ & $-1.5$  &    6437  & 5700 - 7200 & 2.2-2.9  \\
\hline
 \end{tabular}
\end{table*}

If the calibration of Fernley et al. (1998) is preferred, ${\rm [Fe/H]} = -0.195
\Delta S- 0.13$, then the [Fe/H] values are $-1.62 \pm 0.27$, $-0.45 \pm 0.21$ and
$-1.32 \pm 0.29$ for RU Psc, SS Psc and TU UMa respectively. For RU Psc and TU UMa 
the [Fe/H] values reported by Fernley et al. (1998) are $-1.75 \pm 0.15$  and $-1.51
\pm 0.15$. These values were calculated from the $\Delta S$ parameter and the relation
${\rm [Fe/H]} = -0.195 \Delta S- 0.13$. The small differences in [Fe/H] are due to the
use of slightly different values of $\Delta S$.

Therefore, the agreement between the [Fe/H] values from the Fourier light curve 
decomposition (Table \ref{tab:ranges}) and the spectroscopic estimates
from the $\Delta S$ parameter for RU Psc and TU UMa are, within the uncertainties,
very good. 
 
The  $\Delta S$ for SS Psc indicates that the star is much more metal rich than the
RRL stars RU Psc and TU UMa and confirms the comments of McNamara \&
Redcorn (1977) that the star is metal-strong as its spectrum is rich of
metallic lines and its $m_1$ index is large. The mean $m_1$ index in our data
for SS Psc (0.142) is in fact larger than in RU Psc (0.091) and TU UMa (0.088),
consistent with its higher metallicity.

\section{$T_{\rm eff}$ and $\log(g)$ from theoretical grids}

Once the reddening has been inferred, and taking advantage of the simultaneity in the
acquisition of the data in the different color indices, we have plotted our data on
the $(b-y)_o$-$c_1{_o}$ plane along with the theoretical grids calculated by
Lester, Gray \& Kurucz (1986). The $(b-y)_o$-$c_1{_o}$ plane was chosen because loci
of different values of $T_{\rm eff}$ and $log(g)$ are clearly separated. This would
provide good estimates of the mean $T_{\rm eff}$ and $\log(g)$ and the range of
variation along the pulsation cycle.
{\bf The grids are available for a large range of chemical compositions. For each star
we
selected the grid with an [Fe/H] value close to the corresponding [Fe/H]$_{\rm ZW}$
and [Fe/H]($\Delta S$) values in Table \ref{tab:ranges}.}

\begin{figure}
\includegraphics[width=7.5 cm,height=12.0 cm]{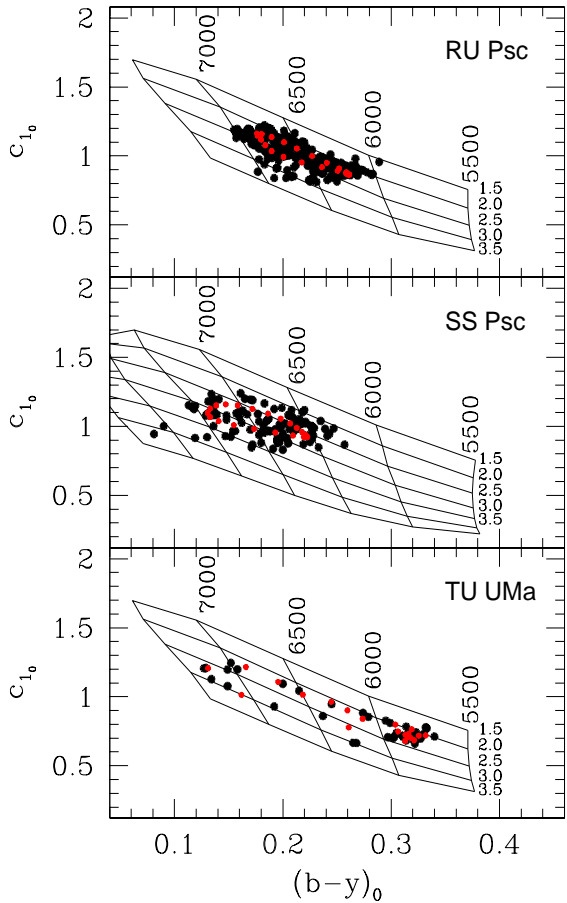}
\caption{Variation of RU Psc, SS Psc and TU UMa on the $(b-y)_o$-$c_1{_o}$ plane
along with the theoretical grids for [M/H] of -1.5, -1.0 and -1.5 respectively, from 
Lester et al. (1986). Nearly vertical
curves are of constant $T_{\rm eff}$, while nearly horizontal curves are for constant
$\log(g)$. Black circles are individual observations. Red circles show the loop
described by the cycle colour variations calculated by Fourier decomposition of the
colour curves.}
\label{fig:3grids}
\end{figure}

The $(b-y)_o$-$c_1{_o}$ planes for RU Psc, SS Psc and TU UMa are presented in Fig.
\ref{fig:3grids}, and the mean $T_{\rm eff}$ and $log(g)$ and their ranges are
given in Table \ref{tab:ranges}. It is interesting to note that SS Psc, which seems
not to be a standard RRc star, but rather a post main sequence star, probably a dwarf
Cepheid (McNamara \& Redcorn 1977) or HADS (Antonello et al. 1986), is indeed much
more metal rich than its RRL counterparts, and the gravity range is slightly
larger than for the two RRL stars, between 2.5 and 3.2. SS Psc also seems to be hotter
than the two RRL stars by about 500 K.

\section{Summary and Conclusions}

From $uvby-\beta$ data collected in several campaigns reported by Pe\~na et al.
(2012) we have addressed the secular variations of the pulsation period of the RRc
star RU Psc, the RRab star TU UMa and the probable RRc star SS Psc. The data
sets for each star span between 3 and 6 years and have been used to provide a
few new times of maximum light. In all cases the data fully covered the pulsation 
cycle and allow the use of the Fourier approach and the loop on the
$(b-y)_o$-$c_1{_o}$ to calculate [Fe/H], $T_{\rm eff}$ and $\log(g)$. 

The three sample stars are subject of secular period instabilities. In the case of RU
Psc a systematic period change cannot be seen, but rather an apparently irregular
period oscillation which causes the observed dispersion in the maxima of the light
curve produced by a data set obtained over 6 years. We found no traces of a
significant secondary period and conclude that the star is not an RRd or double mode
pulsator.
For SS Psc O-C diagram strongly suggests an abrupt period change some time around JD
244 5000 but it can also be argued that the residuals are not inconsistent with a
secularly decreasing period. TU Uma exhibits a complex O-C diagram that has been
interpreted by several authors as being produced by a secular pulsation period change
of the RRL star plus the light time effects in a binary system. The inclusion of the
last ten years of times of maximum clearly define the orbital effect.

The reddening in the direction of each sample star has been estimated from the
reddening sky maps of Schlegel et al. (1998). Then, when possible, we used the
colours $m_1, c_1$ and $\beta$ in the calibrations of Crawford (1975) to calculate
$E(b-y)$. Guided by these results and considering the uncertainties we adopted what
seemed to be 'bona fide' reasonable estimates of $E(B-V)$. 

The physical parameters found in this work for the sample stars are summarized in
Table \ref{tab:ranges}. For the two RRL stars RU Psc and TU UMa the [Fe/H] value
obtained from the Fourier and the $\Delta S$ index methods agree within the
uncertainties. The ranges of $T_{\rm eff}$ and $\log(g)$ from the $(b-y)_o$-$c_1{_o}$
colour distribution bracket well the Fourier value for TU UMa while for RU Psc the
Fourier value is towards the higher limit. The gravity ranges for the three stars are
consistent with values for stars in or near the HB.

SS Psc is much more metal rich than the RRL stars RU Psc and TU UMa, as is indicated
by the several metallicity tracers such as the $\Delta S$ and the mean $m_1$ indices
given in this paper and the richness of metallic lines in its spectrum noticed by
McNamara \& Redcorn (1977). The gravity estimated by these authors, 3.29, is a
bit larger than the range along the pulsation cycle estimated in the present work
(2.5-3.2) despite the slightly larger value of $E(b-y)$ adopted here. Both the
metallicity and gravity are comparable to those in other 
well-established RRL stars, {\bf however, its Fourier $\phi_{31}$ highlights the star
as peculiar among RRc stars in globular clusters.} In our opinion it is not possible
to completely rule out the RRL nature of SS Psc with the available information at
present. 

\section*{Acknowledgments}

We are grateful with Joan Miller for proof reading the
manuscript. This work was partially supported by DGAPA-UNAM through
project IN104612. The authors have made extensive use of the
SIMBAD database operated at CDS, Strasbourg, France and the NASA Astrophysics Data
Systems hosted by Harvard-Smithsonian Center for Astrophysics.


\begin{thebibliography}{}

\bibitem{} Agerer, F., Hubscher, J., 1996, IBVS No. 4382

\bibitem{} Alard C., 2000, A\&AS, 144, 363

\bibitem{}Alcock, C., Allsman, R., Alves, D. R., Axelrod, T.,
Becker, A., Bennett, D., Clement, C., Cook, K. H., Drake, A., Freeman, K., and 19
coauthors, 2000, ApJ, 542, 257.

\bibitem{}Antonello, E.,; Broglia, P., Conconi, P., Mantegazza, L., 1986, A\&A,
169,
122

\bibitem{}Arellano Ferro, A., Ar\'evalo, M.J., L\'azaro, C., Rey, M., Bramich,
D.M.,
Giridhar, S., 2004, RevMexAA, 40, 209


\bibitem{}Arellano Ferro, A., Bramich, D. M., Figuera Jaimes, R., Giridhar, S.,
Kuppuswamy, K., 2012, MNRAS, 420, 1333

\bibitem{}Arellano Ferro, A., Figuera Jaimes, R., Giridhar, S., Bramich, D. M.,
Hern\'andez Santisteban, J. V., Kuppuswamy, K., 2011, MNRAS, 416, 2265

\bibitem{}Arellano Ferro A., Giridhar S., Bramich D. M., 2010, MNRAS, 402, 226

\bibitem{}Arellano Ferro, A., Rojas L\'opez, V., Giridhar, S., Bramich, D.M., 2008,
MNRAS, 384, 1444

\bibitem{} Bla$\check{\rm z}$ko, S., 1907, Astron. Nachr., 175, 325

\bibitem{} Bookmeyer, B. B., Fitch, W. S., Lee, T. A., Wisniewski, W. Z., Johnson,
H. L., 1977, RevMexAA, 2, 235

\bibitem{} Bramich D.M., 2008, MNRAS, 386, L77

\bibitem{} Bramich D.M.,Figuera Jaimes, R., Giridhar, S., Arellano Ferro, A., 2011,
MNRAS, 413, 1275.

\bibitem{}Cacciari, C., Corwin, T. M., Carney, B. W., 2005, ApJ, 129, 267

\bibitem{}Cacciari, C., Clementini, G. 2003, in
Stellar Candles for Extragalactic Distance Scale, ed. D. Alloin \& W. Gieren (Berlin:
Springer), 105

\bibitem{}Chaboyer, B. 1999, in Post-Hipparcos Cosmic Candles, ed. A. Heck \& F.
Caputo (Dordrech: Kluwer), 111

\bibitem{}Clementini, G., Gratton, R., Bragaglia, A., Carretta, E., Di Fabrizio,
L.,
Maio, M. 2003, AJ, 125, 1309

\bibitem{}Crawford, D. L., 1975, AJ, 80, 955

\bibitem{}Derekas, A., Kiss, L. L., Udalski, A., Bedding, T.R., Szatm\'ary, K.,
2004,
MNRAS, 354, 821

\bibitem{}Fernley, J., Skillen, I., Carney, B. W., Cacciari, C., Janes, K., 1998a,
A\&A, 330, 515

\bibitem{}Guggenberger, E., Kolenberg, K., Nemec, J. M., and 13 coauthors, 2012,
MNRAS, 424, 649

\bibitem{}Hauck, B., Mermilliod, M., 1998, A\&AS, 129, 43

\bibitem{}Jurcsik, J., 1995 AcA 653, 660

\bibitem{}Jurcsik, J. 1998, A\&A, 333, 571

\bibitem{}Jurcsik, J., Kov\'acs, G. 1996, A\&A, 312, 111

\bibitem{}Kholopov, P. N. (editor) 1987, General Catalogue of Variable stars,
Nauka,
Moscow.

\bibitem{}Kains, N, Bramich, D.M., Figuera Jaimes, R., Arellano Ferro, A., 
Giridhar, S., Kuppuswamy, K., 2012, A\&A in press

\bibitem{}Kiss, L., Szatmary, K., Gal, J., Kaszas, G., 1995, IBVS, 4205, 1

\bibitem{}Kov\'acs, G. 1998, MmSAI 69, 49

\bibitem{}Kov\'acs, G., Walker, A. R., 2001, A\&A 371, 579

\bibitem{}Layden, A. C., 1994, AJ, 108, 1016.

\bibitem{}L\'azaro, C., Arellano Ferro, A., Ar\'evalo, M.J., Bramich, D.M.,
Giridhar, S., Poretti, E., MNRAS, 2006, 372, 69

\bibitem{}Lenz, P., Breger, M., 2004 IAUS 224, 786

\bibitem{}Lester, J. B., Gray, R. O. \& Kurucz, R. I., 1986 ApJ
61, 509

\bibitem{}McNamara, D.H., Redcorn, M.E., PASP, 1977, 89, 61 

\bibitem{}Mendes de Oliveira, C., Nemec, J. M., 1988, PASP, 100, 217

\bibitem{}Moskalik, P., Olech, A., 2008, CoAst, 157, 345)

\bibitem{}Moskalik, P., Poretti, E., 2003, AA, 398,
213

\bibitem{}Moskalik, P., Smolec, R., Kolenberg, K., et al.,  2012 (arXiv1208.4251).

\bibitem{}Morgan, S., Wahl, J. N., Wieckhorts, R. M., 2007
MNRAS, 374, 1421

\bibitem{}Olech, A., Kaluzny, J., Thompson, I. B., Pych, W., Krzemi\'nski, W.,
Shwarzenberg-Czerny, A., 1999, AJ, 118, 442.

\bibitem{}Olech, A., Kaluzny, J., Thompson, I. B., Pych, W., Krzemi\'nski, W.,
Schwarzenberg-Czerny, A., 2001, MNRAS, 321, 421.
	
\bibitem{}Oosterhoff, P. T., 1939, The Observatory, 62, 104

\bibitem{}Pe\~na, J.~H., Arellano Ferro, A., Pe\~na Miller, R., Sareyan, J.~P.,
\'Alvarez, M.\ 2009, RevMexAA, 45, 191 

\bibitem{}Pe\~na, J.H, Figuera Jaimes, R., Chow, M., Pe\~na Miller, R., \'Alvarez,
M.
2012, RevMexAA, 48, 299

\bibitem{}Preston, G. W., 1959, ApJ, 130, 507

\bibitem{}Rodr\'iguez, E., L\'opez de Coca, P. et al., 1994 AAS, 106, 21

\bibitem{}Saha, A., White, R. E., 1990, PASP, 102, 148

\bibitem{}Schlegel, D. J., Finkbeiner, D. P., Davis, M., 1998, ApJ, 500, 525

\bibitem{}Schuster, W.J., Nissen, P. E., 1988, A\&AS, 73, 225

\bibitem{}Simon, N. R., Clement, C. M. 1993 ApJ 410, 526

\bibitem{}Soszy\'nski, I., Udalski, A., Szyma\'nski, M. K., et. al., 2009, AcA, 59,
1

\bibitem{}Suntzeff, N. B., Kraft, R. P., Kinman, T. D. 1994 ApJS, 93, 271

\bibitem{}Szeidl, B., Olah, K., Mizzer, A., 1986, Comm. Kokoly Obs., No.89

\bibitem{}Tremko, J. 1964, Mitt. Stermw. Akad. Wiss., Budapest, No. 55

\bibitem{}Wade, R. A., Donley, J., Fried, R., White, R. E., Saha, A., 1999 AJ
118, 2442

\bibitem{}Zinn, R., West, M. J, 1984 ApJS 55, 45


\end{thebibliography}
\end{document}